\documentstyle[aps,epsf,preprint,graphicx,prl]{revtex}

\def\beq{\begin{equation}}
\def\eeq{\end{equation}}
\def\beqn{\begin{eqnarray}}
\def\eeqn{\end{eqnarray}}

\def\r {{\bf r}}

\begin{document}

\title{Time-dependent multi-orbital mean-field for
 fragmented Bose-Einstein condensates}

\author{O. E. Alon\footnote{E-mail: ofir@tc.pci.uni-heidelberg.de},
 A. I. Streltsov\footnote{E-mail: alexej@tc.pci.uni-heidelberg.de},
and L. S. Cederbaum\footnote{E-mail: lorenz.cederbaum@pci.uni-heidelberg.de}}

\address{Theoretische Chemie, Physikalisch-Chemisches Institut, Universit\"at Heidelberg,\\
Im Neuenheimer Feld 229, D-69120 Heidelberg, Germany}

\maketitle
\begin{abstract}
The evolution of Bose-Einstein condensates is usually 
described by the famous time-dependent Gross-Pitaevskii equation,
which assumes all bosons to reside in a single time-dependent orbital.
In the present work we address the evolution of fragmented condensates,
for which two (or more) orbitals are occupied,
and derive a corresponding time-dependent multi-orbital mean-field theory.
We call our theory TDMF($n$), where $n$ stands for the number of evolving fragments.
Working equations for a general two-body interaction between the bosons
are explicitly presented along with an illustrative numerical example.
\end{abstract}


\pacs{03.75.Kk, 05.30.Jp, 03.65.-w, 03.75.Nt}

\section{Introduction}

The properties and evolution of Bose-Einstein condensates have been extensively
explored in the community by employing the famous time-independent and time-dependent 
Gross-Pitaevskii equation, for reviews see \cite{Leggett,Stringari_book}
and for individual applications, e.g., Refs.~\cite{R1,R2,R3,R4,R5,R6,R7}.

Gross-Pitaevskii theory is an excellent mean-field for weakly-interacting 
bosons whenever a single macroscopic one-particle wavefunction is sufficient to describe the reality.
By definition, Gross-Pitaevskii theory cannot describe fragmentation of condensates
for which two (or more) one-particle wavefunctions are macroscopically occupied.
Recently, we have developed a time-independent multi-orbital mean-field
to describe static properties of condensates \cite{LA_PLA,OAL_PLA},
thus generalizing the (one-orbital) Gross-Pitaevskii mean-field.
For other approaches to fragmentation of Bose-Einstein condensates,
see, e.g., Refs.~\cite{frag1,frag2,frag3}.
Utilizing the multi-orbital mean-field has enabled us to find new phenomena 
associated with fragmentation as well as fermionization of bosonic atoms in traps and optical lattices 
\cite{ALN_PRA,AL_PRA_2005,OAL_lattice,Us_fermionization}.
Motivated by this wealth of phenomena in the stationary case,
it is the purpose of this work to derive a {\it time-dependent}
multi-orbital mean-field theory,
thus generalizing the {\it time-dependent} Gross-Pitaevskii equation.
This will enable one to study 
dynamical properties of fragmented condensates 
evolving in time.

\section{Preliminaries: Time-dependent Gross-Pitaevskii equation from variational principle}

The evolution of $N$ interacting structureless bosons
is governed by the time-dependent Schr\"odinger equation:
\beq\label{MB_ham}
 \hat H \Phi = i \frac{\partial \Phi}{\partial t}, \qquad 
 \hat H(\r_1,\r_2,\ldots,\r_N) =  \sum_{i=1}^{N} \hat h(\r_i) +  \sum_{i>j=1}^N \lambda_0 V(\r_i-\r_j).
\eeq
Here $\hbar=1$, $\r_i$ is the coordinate of the $i$-th boson, 
$\hat h(\r)$ is the one-body Hamiltonian containing kinetic and potential energy terms,
and $\lambda_0 V(\r_i-\r_j)$ describes the pairwise interaction between the $i$-th and $j$-th atoms
where $\lambda_0$ measures the strength of the interparticle interaction.

The time-dependent Schr\"odinger equation (\ref{MB_ham}) cannot be solved analytically, 
expect for a few specific cases only.
Thus, approximations are a must.
The standard (and simplest) mean-field approximation to the 
time-dependent many-body wavefunction $\Phi$
assumes all bosons to reside in a single 
{\it time-dependent} orbital $\phi(\r,t)$,
namely, that the many-body wavefunction is written as
\beq\label{GP_wavefucntion} 
\Phi(\r_1,\r_2,\ldots,\r_N,t) = \phi(\r_1,t) \phi(\r_2,t) \cdots \phi(\r_N,t).   
\eeq
This so-called Hartree approximation, as is well known, 
gives rise to the famous time-dependent Gross-Pitaevskii equation \cite{Leggett,Stringari_book}. 
For our needs, 
it would be instructive to show how the time-dependent Gross-Pitaevskii equation
derives from a variational principle.
To this end we employ the usual functional action,
see, e.g., Refs.~\cite{DF3,DF4},
which reads:
\beq\label{Hartree_functional}
 S = \int dt \left\{ \left<\Phi\left|\hat H - i\frac{\partial}{\partial t} \right|\Phi\right>
 - \mu(t)\left[\left<\phi|\phi\right> - 1 \right]\right\}, 
\eeq
where the time-dependent Lagrange multiplier $\mu(t)$ is introduced to ensure that
$\phi(\r,t)$ remains normalized throughout the propagation.
To evaluate this action we use the ansatz (\ref{GP_wavefucntion}) and find:
\beqn
& & \left<\Phi\left|\hat H - i\frac{\partial}{\partial t} \right|\Phi\right> = 
 N \int \phi^\ast(\r,t) \left[\hat h(\r) - i\frac{\partial}{\partial t}\right] 
\phi(\r,t) d\r  + \nonumber \\
&+& \lambda_0 \frac{N(N-1)}{2} \int \!\! \int \phi^\ast(\r,t) \phi^\ast(\r',t) V(\r-\r')  
\phi(\r,t) \phi(\r',t) d\r d\r'.\ 
\eeqn
By requiring stationarity of the action with respect to variation of $\phi^\ast(\r,t)$,
namely, $\frac{\delta S}{\delta \phi^\ast(\r,t)} = 0$,
we readily obtain the equation of motion:
\beq\label{pre_GPE}
  N\left[\hat h(\r) + \lambda_0 (N-1) \hat J(\r,t) \right] \phi(\r,t)
= i N \dot\phi(\r,t) + \mu(t)\phi(\r,t),
\eeq
where the direct, time-dependent local potential $\hat J(\r,t)$ reads
\beq\label{GP_non_local}
  \hat J(\r,t) =  \int \phi^\ast(\r',t) V(\r-\r') \phi(\r',t) d\r',
\eeq 
and the shorthand notation $\dot\phi\equiv\frac{\partial \phi}{\partial t}$
has been introduced.

Using the constraint that $\phi(\r,t)$ is normalized
we can eliminate the Lagrange multiplier $\mu(t)$ from Eq.~(\ref{pre_GPE}).
By taking the scalar product of $\phi^\ast(\r,t)$ with (\ref{pre_GPE}), 
the resulting $\mu(t)$ takes on the form,
\beq\label{mu}
 \mu(t) =  N  \int \phi^\ast(\r,t) \left[\hat h(\r) - i\frac{\partial}{\partial t}
+ \lambda_0 (N-1) \hat J(\r,t) \right] \phi(\r,t) d\r. 
\eeq 
Substituting Eq.~(\ref{mu}) into Eq.~(\ref{pre_GPE}) and noticing the identity
\beqn\label{GPE_dot_phi_0}
 & & \left[\hat h(\r) - i\frac{\partial}{\partial t} + \lambda_0 (N-1) \hat J(\r,t)\right] \phi(\r,t) - 
 \frac{\mu(t)}{N}\phi(\r,t) = \nonumber \\ 
&=& \left(1 - \phi(\r,t) \int \phi^\ast(\r,t) d\r \right) 
\left[\hat h(\r) - i\frac{\partial}{\partial t} + \lambda_0 (N-1) \hat J(\r,t) \right] \phi(\r,t), \
\eeqn
we arrive immediately at the time-dependent mean-field equation:
\beqn\label{GPE0_dot_phi}
& &  {\mathbf P}  i\dot\phi(\r,t) = {\mathbf P} \left[\hat h(\r) + \lambda_0 (N-1) \hat J(\r,t) \right] \phi(\r,t), 
\nonumber \\
& & \qquad {\mathbf P}=1 - \phi(\r,t) \int \phi^\ast(\r,t) d\r. \
\eeqn
Examining Eq.~(\ref{GPE0_dot_phi}) we see that
eliminating the Lagrange multiplier $\mu(t)$
has emerged as a projection operator ${\mathbf P}$ 
onto the subspace spanned by $\phi(\r,t)$.
This projection appears both on the left- and right-hand-sides of (\ref{GPE0_dot_phi}),
making it a somewhat cumbersome integro-differential non-linear equation.
To simplify its appearance we choose the relation
\beq\label{const}
 \int \phi^\ast(\r,t) \dot\phi(\r,t) d\r = 0
\eeq
at any time. 
This condition is equivalent to making the assignment of the time-dependent phase
$\phi(\r,t) \rightarrow \exp\left\{+\int\left<\phi|\dot\phi\right> dt\right\}\phi(\r,t)$
in Eq.~(\ref{GPE0_dot_phi}).
With this condition, the influence of the projection on the left-hand-side simplifies, 
${\mathbf P}\dot\phi(\r,t)=\dot\phi(\r,t)$, and (\ref{GPE0_dot_phi}) takes on the appealing from:
\beqn\label{GPE_dot_phi}
& &  i \dot\phi(\r,t) = {\mathbf P} \left[\hat h(\r) + \lambda_0 (N-1) \hat J(\r,t) \right] \phi(\r,t), 
\nonumber \\
& & \qquad {\mathbf P}=1 - \phi(\r,t) \int \phi^\ast(\r,t) d\r. \
\eeqn
The ${\mathbf P}$ remaining on the right-hand-side of Eq.~(\ref{GPE_dot_phi}) makes it clear that
the condition (\ref{const}) is indeed satisfied at any time throughout the propagation of $\phi(\r,t)$. 
In practice, the meaning of this condition is that the temporal change
of $\phi(\r,t)$ is always orthogonal to $\phi(\r,t)$ itself,
which can be exploited to maintain accurate propagation results at lower computational costs.   
Additionally with condition (\ref{const}),
 $\mu(t)$ in Eq.~(\ref{mu}) takes now 
an appealing form which can be interpreted as ($N$ times) the time-dependent chemical potential of the condensate. 

To complete this section we would like to show that the three equations,
Eq.~(\ref{GPE0_dot_phi}), Eq.~(\ref{GPE_dot_phi})
and the equation obtained by 'omitting' the projector ${\mathbf P}$ completely
are fully equivalent to each other.
To this end, we notice that the Lagrange multiplier $\mu(t)$ 
can be 'absorbed' into $\phi(\r,t)$ by making the assignment 
$\phi(\r,t) \rightarrow \exp\left\{i/N \int \mu(t) dt \right\}\phi(\r,t)$.
This immediately results in the time-dependent 
Hartree mean-field equation
\beq\label{GPE_simple}
  i \dot\phi(\r,t) = \left[\hat h(\r) + \lambda_0 (N-1) \hat J(\r,t) \right] \phi(\r,t). 
\eeq
In other words, in the case of a single orbital $\phi$,
one can either use the form (\ref{GPE0_dot_phi}), the form (\ref{GPE_dot_phi}) 
or the form (\ref{GPE_simple}).

Finally, to arrive at the time-dependent Gross-Pitaevskii equation 
in its usual appearance we take $\lambda_0V(\r-\r')=\lambda_0\delta(\r-\r')$ 
and set $\lambda_0$ proportional to the s-wave scattering length in any of the 
above equivalent forms (\ref{GPE0_dot_phi}), (\ref{GPE_dot_phi}) or (\ref{GPE_simple}).
Consequently, the direct potential (\ref{GP_non_local}) simplifies,
$\hat J(\r,t)=\left|\phi(\r,t)\right|^2$, leading to the familiar 
result \cite{Leggett,Stringari_book}:
\beq\label{TDGPE}
 i \dot\phi(\r,t) = \left[ \hat h(\r) + \lambda_0 (N-1) \left|\phi(\r,t)\right|^2 \right] \phi(\r,t).  
\eeq

In what follows we generalize the time-dependent Gross-Pitaevskii equation, 
which is only valid for non-fragmented condensates, 
by constructing a multi-orbital time-dependent mean-field for fragmented condensates.
We will generalize the forms (\ref{GPE0_dot_phi}) and (\ref{GPE_dot_phi}) 
of the time-dependent Gross-Pitaevskii equation
where in the latter case constraints such as 
(\ref{const}) would make the interpretation of the 
results as well as the numerical 
implementation particularly useful.

\section{Time-dependent multi-orbital mean-field equations for bosons}

Generalizing the stationary multi-orbital mean-field ansatz \cite{LA_PLA,OAL_PLA}, 
the most general time-dependent mean-field {\it wavefunction} for $N$ interacting bosons is
the {\it single} configuration time-dependent wavefunction 
\beq\label{BMF_wavefucntion}
 \Phi(\r_1,\r_2,\ldots,\r_N,t) = \hat{\mathcal S}
 \phi_1(\r_1,t) \phi_2(\r_2,t) \cdots \phi_N(\r_N,t),
\eeq
where $\hat{\mathcal S}$ is the symmetrization operator.
Of course, for bosons not all $\phi_k(\r,t)$ have to be different functions.
For instance, in the Hartree (or Gross-Pitaevskii) ansatz 
(\ref{GP_wavefucntion}) all $\phi_k(\r,t)$ are taken to be equal to one another.
Generally, however, 
we may take $n_1$ bosons to reside in $\phi_1(\r,t)$,
$n_2$ bosons to reside in a different orbital $\phi_2(\r,t)$, and so on,
as long as the set of occupations $\{n_k\}$ satisfies 
$n_1+n_2+\ldots+n_{orb}=N$
and, of course, 
the number $n_{orb}$ of different orbitals satisfies $1 \le n_{orb} \le N$.
What is important to note is that the different orbitals $\phi_k(\r,t)$
within ansatz (\ref{BMF_wavefucntion}) 
are propagated variationally in time
and remain {\it normalized} and {\it orthogonal} to one another at any time.

To proceed for any set of $n_{orb}$ orbitals and a corresponding set 
of occupations $\{n_k\}$,
we substitute the ansatz (\ref{BMF_wavefucntion}) for $\Phi$ 
into the usual functional action which now reads:
\beq\label{BMF_functional}
 S = \int dt \left\{ \left<\Phi\left|\hat H - i\frac{\partial}{\partial t} \right|\Phi\right>
 - \sum_{k,j}^{n_{orb}} \mu_{kj}(t)\left[\left<\phi_k|\phi_j\right> - \delta_{kj} \right]\right\}, 
\eeq
where the time-dependent Lagrange multipliers $\mu_{kj}(t)$ are introduced to ensure that
the time-dependent orbitals $\phi_k(\r,t)$ remain orthonormal throughout the propagation.
The expectation value in the action takes on the appearance
\beqn
& & \left<\Phi\left|\hat H - i\frac{\partial}{\partial t} \right|\Phi\right> = \nonumber \\
& & \qquad = \sum_{k}^{n_{orb}} n_k \left[h_{kk} -\left(i\frac{\partial}{\partial t}\right)_{kk} + 
 \lambda_0 \frac{n_k-1}{2} V_{kkkk} + 
  \frac{1}{2}\sum_{l\ne k}^{n_{orb}} \lambda_0 n_l V_{kl[kl]} \right],  \
\eeqn
where the one- and two-body matrix elements are given by
\beqn
 h_{kj} &=& \int \phi_k^\ast(\r) h(\r) \phi_j(\r) d\r, \nonumber \\
 \left(i\frac{\partial}{\partial t}\right)_{kj} &=& 
i\int \phi_k^\ast(\r) \dot \phi_j(\r) d\r, \nonumber \\
V_{kjql} &=& \int \!\! \int \phi_k^\ast(\r) \phi_j^\ast(\r') V(\r-\r')  
 \phi_q(\r) \phi_l(\r') d\r d\r', \nonumber \\
 V_{kj[ql]} &=& V_{kjql} + V_{kjlq}. 
\eeqn
Note the plus sign in $V_{kj[ql]}$ which is due to the symmetrization operator.

By requiring stationarity of the action 
with respect to variation of all $\phi^\ast_k(\r,t)$, 
namely, $\frac{\delta S}{\delta \phi^\ast_k(\r,t)} = 0, k=1,\ldots,n_{orb}$, 
we obtain a set of $n_{orb}$
equations of motion 
for the multi-orbital mean-field ansatz (\ref{BMF_wavefucntion}):
\beqn\label{BMF_equations_general} 
 & & n_k\left[\hat h -i\frac{\partial}{\partial t} 
+ \lambda_0 (n_k-1) \hat J_{k} + 
  \sum_{l\ne k}^{n_{orb}} \lambda_0 n_l \left(\hat J_l + \hat K_l \right) \right] 
\left|\phi_k\right> = \nonumber \\
& & \qquad = \sum_{j}^{n_{orb}} \mu_{kj}(t) \left|\phi_j\right>, 
 \qquad k=1,\ldots,n_{orb}, \
\eeqn
where the direct (local) and exchange (non-local)
 time-dependent potentials are given by 
\beqn\label{one_body_pot}
& &  \hat J_l(\r,t) = \int \phi_l^\ast(\r',t) V(\r-\r')
 \phi_l(\r',t) d\r', \nonumber \\
& &  \hat K_l(\r,t) = \int \phi_l^\ast(\r',t) V(\r-\r') {\mathcal P}_{\r\r'} \phi_l(\r',t) d\r', \
\eeqn
and ${\mathcal P}_{\r\r'}$ permutes the $\r$ and $\r'$ coordinates of two bosons
appearing to the right of it.

Using the constraints that the $\phi_k(\r,t)$ are orthonormal functions
we can eliminate the Lagrange multipliers $\mu_{kj}(t)$ from Eq.~(\ref{BMF_equations_general}).
By taking the scalar product of $\left<\phi_j\right|$ 
with (\ref{BMF_equations_general}), 
the resulting $\mu_{kj}(t)$ take on the form,
\beq\label{BMF_mu}
 \mu_{kj}(t) =  n_k \left[ h_{jk} -  \left(i\frac{\partial}{\partial t}\right)_{jk} + \lambda_0(n_k-1)V_{jkkk} +
 \sum_{l\ne k}^{n_{orb}} \lambda_0 n_l V_{jl[kl]}\right].
\eeq 
Substituting Eq.~(\ref{BMF_mu}) into Eq.~(\ref{BMF_equations_general}) 
and noticing the identities 
\beqn\label{TDMF_equ_0}
& & \!\!\!\!\!\!\!\!\!\!\!\!\! \left[\hat h -i\frac{\partial}{\partial t} + \lambda_0 (n_k-1) \hat J_k + 
 \sum_{l\ne k}^{n_{orb}} \lambda_0 n_l \left(\hat J_l + \hat K_l\right) \right] \left|\phi_k\right>   
 - \frac{1}{n_k} \sum_{j}^{n_{orb}} \mu_{kj}(t)\left|\phi_j\right> = \nonumber \\
& & \!\!\!\!\!\!\!\!\!\!\!\!\! \left(1-\sum_{j=1}^{n_{orb}}\left|\phi_j\left>\right<\phi_j\right|\right)
\left[\hat h -i\frac{\partial}{\partial t} + \lambda_0 (n_k-1) \hat J_k + 
 \sum_{l\ne k}^{n_{orb}} \lambda_0 n_l \left(\hat J_l +\hat K_l\right)\right] \left|\phi_k\right> \ 
\eeqn
for all $k$,
we arrive immediately at the time-dependent multi-orbital mean-field equations, $k=1,\ldots,n_{orb}$:
\beqn\label{TDMF0_equ}
& & 
{\mathbf P} i\left|\dot\phi_k\right> = {\mathbf P} \left[\hat h + \lambda_0 (n_k-1) \hat J_k + 
 \sum_{l\ne k}^{n_{orb}} \lambda_0 n_l \left(\hat J_l +\hat K_l\right)\right] \left|\phi_k\right>, 
\nonumber \\
& & \qquad {\mathbf P} = 1-\sum_{j=1}^{n_{orb}}\left|\phi_j\left>\right<\phi_j\right|. \
\eeqn
Examining Eq.~(\ref{TDMF0_equ}) we see that
eliminating the Lagrange multipliers $\mu_{kj}(t)$
has emerged as a projection operator ${\mathbf P}$ 
onto the subspace spanned by the $\phi_k(\r,t)$. 
This projection appears both on the left- and right-hand-sides of (\ref{TDMF0_equ}),
making it a coupled system of integro-differential non-linear equations.

The equation set (\ref{TDMF0_equ}) is the time-dependent multi-orbital mean-field 
which generalizes the one-orbital time-dependent mean-field equation (\ref{GPE0_dot_phi}).
We call our theory in short TDMF($n$), where $n=n_{orb}$ stands for the number of evolving fragments. 
Similarly to the stationary case \cite{LA_PLA,OAL_PLA},
the multi-orbital time-dependent theory boils down to the standard 
time-dependent mean-field for $n_{orb}=1$. 
In contrast to the stationary case, 
where the occupations $n_k$ of the fragments are also determined variationally,
the $n_k$ are determined by the initial condition, say at $t=0$,
and remain unaltered during the propagation.
To enable the variation of the occupations in time,
a much more involved many-body theory is needed.
Such a theory is possible \cite{MCus}.

The multi-orbital coupled system (\ref{TDMF0_equ}) is more involved than its one-orbital
predecessor Eq.~(\ref{GPE0_dot_phi}). 
Naturally and similarly to the previous section, 
we would like to make its appearance simpler.
We can choose the relations 
\beq\label{const_mu_kk}
  \left<\phi_k|\dot\phi_k\right>= 0, \ \ k=1,\ldots,n_{orb}
\eeq
at any time,
which is equivalent  
to making the assignments of the time-dependent phases
$\phi_k(\r,t) \rightarrow \exp\left\{+\int\left<\phi_k|\dot\phi_k\right> dt\right\}\phi_k(\r,t)$
in Eq.~(\ref{TDMF0_equ}).
With these conditions, the influence of the projectors on the left-hand-side slightly simplifies, 
${\mathbf P}\left|\dot\phi_k\right>={\mathbf P_k}\left|\dot\phi_k\right>$, 
where ${\mathbf P_k} = 1-\sum_{j\ne k}^{n_{orb}}\left|\phi_j\left>\right<\phi_j\right|$.
Additionally, the diagonal Lagrange multiplier $\mu_{kk}(t)$, see Eq.~(\ref{BMF_mu}),
can now be interpreted as ($n_k$ times) the time-dependent chemical potential 
of the $k$-th fragment, thus generalizing the time-independent quantities \cite{LA_PLA,AL_PRA_2005}.
However, since the off-diagonal terms 
$\left<\phi_k|\dot\phi_j\right>, k \ne j$
can {\it a priori} not be absorbed in a similar manner 
by time-dependent phases, the system of coupled TDMF($n$) equations, 
combining Eqs.~(\ref{TDMF0_equ}) and (\ref{const_mu_kk}), 
remains involved. 

To proceed let us examine a few properties of 
the time-dependent multi-orbital mean-field for bosons, TDMF($n$).
First, it is clear that an initially normalized permanent $\Phi$ 
remains normalized throughout the propagation since the orbitals $\phi_k(\r,t)$ 
remain orthonormal functions at any time.
Second, the expectation value of the Hamiltonian with 
respect to $\Phi$ should be constant if the many-body Hamiltonian $\hat H$ is time-independent.
To examine this, we calculate the time derivative of $\left<\Phi\left|H\right|\Phi\right>$.
Taking the scalar product of $\dot \phi_k(\r,t)$ with Eq.~(\ref{BMF_equations_general}),
summing up over $k$ and adding the result to its complex conjugate 
we readily obtain:
\beqn\label{energy}
 \frac{d}{dt}\left<\Phi\left|\hat H\right|\Phi\right> &=& 
\sum_{k}^{n_{orb}} n_k \bigg[h_{\dot kk} + h_{k\dot k} +  
 \lambda_0 (n_k-1) \left(V_{\dot kkkk} + V_{kkk\dot k}\right) + \nonumber \\
&+&  \sum_{l\ne k}^{n_{orb}} \lambda_0 n_l \left(V_{\dot kl[kl]} + V_{kl[\dot kl]} \right)\bigg] =  \\
 & & \!\!\!\!\!\!\!\!\!\!\!\!\!\!\!\!\!\!\!\!  
 = \sum_{k,j}^{n_{orb}} \left[\mu_{kj}(t) \left<\dot\phi_k|\phi_j\right> +
        \mu^\ast_{jk}(t) \left<\phi_k|\dot\phi_j\right>\right]. \nonumber \    
\eeqn
In (\ref{energy}) we use the shorthand notation for the index $\dot k$
to indicate expectation values with respect to $\dot \phi_k(\r,t)$,
e.g., $h_{\dot kk}=\left<\dot\phi_k|\hat h|\phi_k\right>$, etc.
From Eq.~(\ref{energy}) we see that the energy would be conserved if the
matrix of Lagrange multiplier is Hermitian, $\mu_{kj}(t)=\mu^\ast_{jk}(t)$,
simply because
\beq
 \left<\dot\phi_k|\phi_j\right> + \left<\phi_k|\dot\phi_j\right> = 0, \ \ k,j=1,\ldots,n_{orb}
\eeq  
hold due to the orthonormality constraints.
Whether throughout the time propagation of Eq.~(\ref{TDMF0_equ}) the matrix 
of Lagrange multipliers $\mu_{kj}(t)$ can be kept Hermitian at all times 
is a matter of further studies.
There is an alternative to this Hermiticity condition, 
in the form of solutions to Eq.~(\ref{TDMF0_equ}) which obey the constraints \cite{CPL} 
\beq\label{BMF_const}
  \left<\phi_k|\dot\phi_j\right> = 0, \ \ k,j=1,\ldots,n_{orb} 
\eeq
at any time. 
When these additional constraints are explicitly invoked,
the energy is readily conserved, see Eq.~(\ref{energy}).
The influence of the projection on $\left|\dot\phi_k\right>$
simplifies considerably, 
${\mathbf P}\left|\dot\phi_k\right>=\left|\dot\phi_k\right>$,
and thus (\ref{TDMF0_equ}) takes on the appealing from, $k=1,\ldots,n_{orb}$:
\beqn\label{TDMF_equ}
& & 
 i\left|\dot\phi_k\right> = {\mathbf P} \left[\hat h + \lambda_0 (n_k-1) \hat J_k + 
 \sum_{l\ne k}^{n_{orb}} \lambda_0 n_l \left(\hat J_l +\hat K_l\right)\right] \left|\phi_k\right>, 
\nonumber \\
& & \qquad {\mathbf P} = \left(1-\sum_{j=1}^{n_{orb}}\left|\phi_j\left>\right<\phi_j\right|\right). \
\eeqn
The ${\mathbf P}$ remaining on the right-hand-side of Eq.~(\ref{TDMF_equ}) 
makes it clear that
the constraints (\ref{BMF_const}) are indeed satisfied at any time throughout the propagation of $\phi_k(\r,t)$. 
In practice, the meaning of these constraints is that the 
temporal changes of the $\phi_k(\r,t)$ are always orthogonal to the $\phi_k(\r,t)$ themselves,
a property which makes the time propagation of Eq.~(\ref{TDMF_equ}) robust and stable and 
which can thus be exploited to maintain accurate propagation results at lower computational costs.   

To complete our derivation we would like to pose the question whether,
similarly to the property of the time-dependent Hartree mean-field
discussed in the previous section, see Eq.~(\ref{GPE_simple}),
the projector ${\mathbf P}$ may totally be 'omitted'
in the TDMF($n$) theory.
Examining Eq.~(\ref{BMF_equations_general}) 
we note that the diagonal Lagrange multipliers $\mu_{kk}(t)$
can indeed be 'absorbed' into the orbitals $\phi_k(\r,t)$
by making the assignments
$\phi_k(\r,t) \rightarrow \exp\left\{i/n_k \int \mu_{kk}(t) dt \right\}\phi_k(\r,t)$.
However, in contrast to this situation
the off diagonal Lagrange multipliers
$\mu_{kj}(t)$, $k\ne j$,
which ensures orthogonality of the orbitals at any time, 
cannot in general be 'absorbed' into the evolution of the $\phi_k(\r,t)$,
namely that the form (\ref{GPE_simple}) does not generally exist in the TDMF($n$) theory. 
Therefore and in view of all the above,
it is the version (\ref{TDMF_equ}) of the TDMF($n$) 
which will be implemented and employed below in our numerical example.

Finally, we mention that the presently developed 
time-dependent multi-orbital theory and its stationary version
are intimately connected via the so-called imaginary-time 
propagation technique.
It is well known that the ground state of the  
stationary Gross-Pitaevskii equation 
can be obtained by propagating in imaginary time 
the time-dependent Gross-Pitaevskii equation, 
see, e.g., \cite{R5}.
In our case, by setting $t \to -it$, 
the left-hand-sides of (\ref{TDMF0_equ}) and (\ref{TDMF_equ}) decay to zero in time.
Then, by using the identity (\ref{TDMF_equ_0}) 
and the expression (\ref{BMF_mu}) for the Lagrange multipliers,
the right-hand-sides of (\ref{TDMF0_equ}) and (\ref{TDMF_equ}) 
boil down to the stationary multi-orbital mean-field theory \cite{LA_PLA,OAL_PLA}.
This establishes a natural connection between the
time-dependent and time-independent 
multi-orbital mean-fields for bosons.

\section{Illustrative numerical example}

Having derived the time-dependent multi-orbital mean-field  TDMF($n$)
we wish to apply it to a specific problem of evolution of fragmented condensates.
Of course, we have to chose a specific
shape for the interparticle interaction
and we do so by taking the popular contact interaction,
$\lambda_0 V(\r_i-\r_j) = \lambda_0 \delta(\r_i-\r_j)$,
see Refs.~\cite{Leggett,Stringari_book}. 
The resulting set of $n_{orb}$ equations (\ref{TDMF_equ}) simplifies considerably
and is given by, $k=1,\ldots,n_{orb}$:
\beqn\label{TDMF_equations_contact} 
 & & i\left|\dot\phi_k\right> = 
{\mathbf P} \left[\hat h + \lambda_0(n_k-1)|\phi_k|^2 + 
\sum_{l\ne k}^{n_{orb}} 2\lambda_0 n_l |\phi_l|^2\right] \left|\phi_k\right>, 
  \nonumber \\
 & & \qquad {\mathbf P} = \left(1-\sum_{j=1}^{n_{orb}}\left|\phi_j\left>\right<\phi_j\right|\right).
\eeqn
The factor of $2$ in the last term of the square braces
comes from the fact that {\it both} the exchange and direct potentials are
now local potentials which are {\it equal} to one another, 
namely, $\hat J_l(\r,t)=\hat K_l(\r,t)=|\phi_l(\r,t)|^2$.
Eq.~(\ref{TDMF_equations_contact}) 
is the generalization of the time-dependent Gross-Pitaevskii
equation (\ref{TDGPE}) and will be applied below for a one-dimensional problem
of a two-fold fragmented condensate evolving in time,
i.e., for $n_{orb}=2$.

As an illustrative numerical example we consider $N=1000$ interacting bosons in a double-well potential.
The numerical implementation of TDMF($2$) employs 
the discrete variable representation (DVR) method \cite{DVR} 
and Adams-Bashforth-Moulton predictor-corrector integrator \cite{abm}.
It is convenient to work in dimensionless units
in which the one-body Hamiltonian reads $\hat h(x)=-\frac{1}{2}\frac{\partial^2}{\partial x^2} + V(x)$.
As the trapping potential we chose the symmetric potential $V(x)=0.05x^2+10/\sqrt{2\pi}e^{-x^2/8}$.
The interparticle interaction strength is $\lambda_0=0.1$.
Initially, the system is prepared in the symmetric double-well $V(x)$
in a two-fold fragmented (or, Fock) state $\Phi=\left|n_1,n_2\right>$, 
with $n_1=n_2=N/2$ bosons per fragment. 
The two orthonormal orbitals $\phi_1(x,t=0)$ and $\phi_2(x,t=0)$
in which the fragments initially reside are localized in the left and right wells of $V(x)$. 
At time $t=0$, the symmetric potential trap $V(x)$ is suddenly translated to $V(x+2)$
and the system is subsequently allowed to propagate in time in the new 
shifted potential $V(x+2)$ according to the TDMF equation (\ref{TDMF_equations_contact}).
Since the system is not in a stationary state any more,
the density
\beq\label{density}
\rho(x,t)=\sum_k^{n_{orb}} n_k \left|\phi_k(x,t)\right|^2=
  \frac{N}{2}\left[\left|\phi_1(x,t)\right|^2+\left|\phi_2(x,t)\right|^2\right]
\eeq  
evolves in time.

In Fig.~1 we record four snapshots of the time-dependent density $\rho(x,t)$. 
As can be seen in the figure, the density profile as a whole oscillates from side to side,
which is expected from this scenario where the trap has been shifted at $t=0$.
Atop this behavior of $\rho(x,t)$, 
we observe that the initial sharp dip of the 
fragmented-state density at $t=0$ nearly vanishes at $t=0.9$,
then it is restored at $t=5.5$, and again it nearly vanishes at $t=6.7$.
Furthermore, a few density wiggles which vary in time  
accommodate $\rho(x,t)$ in addition to the above features,
all demonstrating that the time-evolution of fragmented 
condensates is both an intricate and appealing dynamical problem
which awaits and invites ample investigations. 

\section{Summary}

The evolution of Bose-Einstein condensates has been extensively studied
by the well-known time-dependent Gross-Pitaevskii equation,
which assumes all bosons to reside in the same time-dependent orbital.
In this work we addressed the evolution of fragmented condensates,
for which two (or more) orbitals are occupied,
and derived a corresponding time-dependent multi-orbital mean-field theory,
thus generalizing the (one-orbital) time-dependent Gross-Pitaevskii mean-field.
We call our theory TDMF($n$), where 
$n=n_{orb}$ stands for the number of evolving fragments (orbitals).

In the TDMF, the ansatz for the many-body wavefunction 
is taken as a permanent made of $n_{orb}$ time-dependent orthonormal orbitals,
whose occupations are determined, say, by the initial conditions.
The evolution of the many-body wavefunction 
is then determined variationally by utilizing the usual functional action.
This results in a set of $n_{orb}$ coupled time-dependent equations 
for the evolution of the $n_{orb}$ orbitals (fragments) in time.
Working equations for a general two-body interaction between the bosons are explicitly presented. 
By considering their propagation in imaginary time,
TDMF is shown to naturally relate to the recently developed and successfully employed 
{\it stationary} multi-orbital mean-field theory.

Finally, an illustrative numerical example of 
the evolution of a two-fold fragmented condensate in a one-dimensional 
double-well trap is provided, demonstrating an intricate dynamics
of the density as time progresses. 
This is the tip of the iceberg of dynamical properties of fragmented Bose-Einstein condensates
which await many more investigations. 

\acknowledgments
  
\noindent
We thank Hans-Dieter Meyer for helpful discussions 
and making available the numerical integrator.


\newpage

\begin{figure}
\includegraphics[width=10cm,angle=-90]{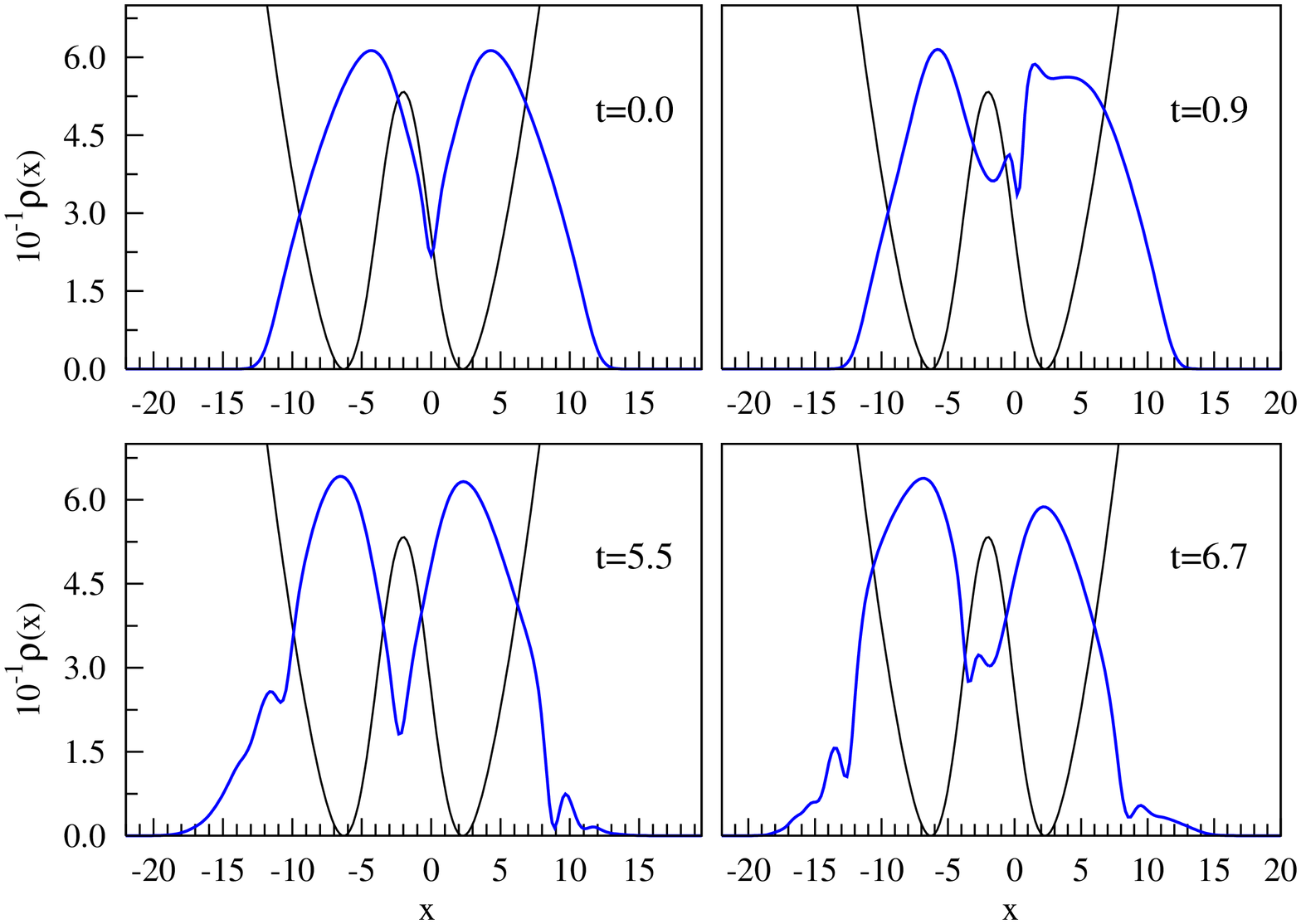}
\vglue 1.0 truecm
\caption{(Color online) Time evolution of fragmented condensate 
made of $N=1000$ bosons in a double-well potential.
A two-fold fragmented state $\Phi=\left|N/2,N/2\right>$ is initially 
prepared in a symmetric double-well trap $V(x)$. 
At time $t=0$, the trap is suddenly translated a bit to the left, 
namely, $V(x) \to V(x+2)$, and the fragmented condensate
is allowed to propagate in time according to the TDMF equation (\ref{TDMF_equations_contact}).
Shown are four time snapshots of the density (blue curves), Eq.~(\ref{density}).
It exhibits an intricate record of 
overall oscillations from side to side,
evolution of the initial, central dip,
and appearance of additional density wiggles 
which varies in time.
For convenience, the trap potential $V(x+2)$ (black curves) is shown, 
scaled by $2$ and shifted vertically 
such that $V=0$ at its minima.
} 
\label{fig1}
\end{figure}

\end{document}